\documentclass[twocolumn,aps,floatfix,superscriptaddress,showpacs,citeautoscript]{revtex4}
\usepackage{graphicx}
\usepackage{amssymb}
\usepackage{pifont}

\begin{document}

\title{Suppression of spin-pumping by a MgO tunnel-barrier}

\author{O.~Mosendz}
\email{mosendz@anl.gov} \affiliation{Materials Science Division,
Argonne National Laboratory, Argonne, IL 60439, USA}

\author{J.~E.~Pearson}
\affiliation{Materials Science Division, Argonne National
Laboratory, Argonne, IL 60439, USA}

\author{F.~Y.~Fradin}
\affiliation{Materials Science Division, Argonne National
Laboratory, Argonne, IL 60439, USA}

\author{S.~D.~Bader}
\affiliation{Materials Science Division, Argonne National
Laboratory, Argonne, IL 60439, USA} \affiliation{Center for
Nanoscale Materials, Argonne National Laboratory, Argonne, IL 60439,
USA}

\author{A.~Hoffmann}
\affiliation{Materials Science Division, Argonne National
Laboratory, Argonne, IL 60439, USA} \affiliation{Center for
Nanoscale Materials, Argonne National Laboratory, Argonne, IL 60439,
USA}



\date{\today}

\begin{abstract}
Spin-pumping generates pure spin currents in normal metals at the
ferromagnet (F)/normal metal (N) interface. The efficiency of
spin-pumping is given by the spin mixing conductance, which depends
on N and the F/N interface. We directly study the spin-pumping
through an MgO tunnel-barrier using the inverse spin Hall effect,
which couples spin and charge currents and provides a direct
electrical detection of spin currents in the normal metal. We find
that spin-pumping is suppressed by the tunnel-barrier, which is
contrary to recent studies that suggest that the spin mixing
conductance can be enhanced by a tunnel-barrier inserted at the
interface.
\end{abstract}



\maketitle


Developments in spintronics provide new insights in the physics of
spin related phenomena. One new research direction is to explore
pure spin currents, which are independent of charge
currents~\cite{Chappert-NP2008,Hoffmann-PSSC2007}. Pure spin
currents can be generated via a spin-polarized charge current
injection from ferromagnets
\cite{Johnson-PRL1985,Jedema-Nature2001}, spin Hall effects
\cite{Dyakonov-PLA,Hirsch}, or spin-pumping
\cite{Heinrich-PRL2003,Brataas-PRL2002}. The last differs from the
other two mechanisms, since it uses magnetization dynamics rather
than electric charge currents for the pure spin current generation.
The spin current in this case is generated by a precessing
magnetization in the ferromagnetic layer at the ferromagnet
(F)/normal metal (N) interface. The precessing magnetization at the
F/N interface acts as a peristaltic spin-pump, which creates a
dynamic spin accumulation in the normal metal diffusing away from
the interface. Non-local effects in the magnetization dynamics due
to spin-pumping were studied extensively
\cite{Woltersdorf-PRL2007,Mosendz-PRB09,Kardasz-JAP2008,Mosendz-JAP08}
and revealed a new coupling mechanism between ferromagnets separated
by a non-magnetic material. Furthermore, studies, which used
spin-pumping as a spin current generator enabled quantification of
spin Hall effects in normal metals
\cite{Mosendz-PRL09,Saitoh-APL06}. For practical applications it
would be useful to increase the efficiency of spin-pumping.  To this
end, recent work by Moriyama \textit{et al.}\ \cite{Moriyama-PRL08}
suggests that spin-pumping may be significantly enhanced by an
insulating (I) tunnel-barrier inserted at the interface between a
ferromagnet and normal metal. This result contrasts with theoretical
predictions~\cite{Brataas-PRB02}, which suggests that sizable
spin-pumping requires a transparent interface between the
ferromagnet and normal metal.
\begin{figure}
  \includegraphics[width=6.5cm]{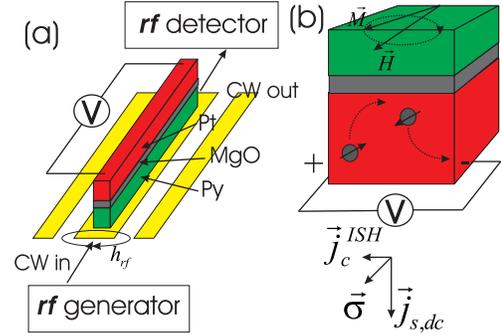}
  \caption{(Color online) Experimental setup.  (a) Schematic of the Py/MgO/Pt structure integrated into the coplanar waveguide. (b) ISHE detection schematic. The precessing magnetization in the ferromagnet generates a spin current in the adjacent normal metal. This spin current leads to a charge current in the normal metal due to a preferential scattering direction for electrons carrying the pumped spin accumulation. The charge current is orthogonal to the spin current direction and the spin polarization. A voltage difference due to the ISHE is measured between the ends of Pt/MgO/Py strip. Note that orientation in (a) and (b) is changed for illustrative purposes.} %
  \label{setup1}%
\end{figure}

To resolve this problem a detection scheme sensitive to the pure
spin currents in the normal metal is needed. Inverse spin Hall
effect (ISHE) measurements \cite{Mosendz-PRL09,Saitoh-APL06} are a
good approach to detect any enhancement of spin pumping across the
F/I/N structure, since they are directly sensitive to pure spin
currents and unaffected by other spurious voltages across the
tunnel-barrier. The {\em dc} part of the spin current in the normal
metal generated by the spin-pumping, gives rise to a transverse
charge current $\vec{j}_{c}^{ISH}= \gamma (2e/\hbar)[\vec{j}_{s,dc}
\times \vec{\sigma}]$ via the ISHE [see Fig.~\ref{setup1}(b)] where
$\gamma$ is the materials-specific spin Hall angle, $\vec{j}_{s,dc}$
is the spin current and $\vec{\sigma}$ is the spin polarization. Any
enhancement of spin-pumping will manifest itself as an increased
voltage due to the ISHE. Thus comparing $\vec{j}_{c}^{ISH}$ for two
structures: (i) F/N and (ii) F/I/N enables us to investigate the
spin-pumping strength in the case of an insulating tunnel-barrier
compared to a transparent interface.  As we show in this Letter,
spin-pumping is actually suppressed by the tunneling barrier.

We integrated different F/N heterostructures into coplanar
waveguides with additional leads for {\em dc} voltage measurements
along the sample. A schematic is shown in Fig.~\ref{setup1}(a) for a
15Ni$_{80}$Fe$_{20}$ (Py)/3MgO/15Pt heterostructure, where integers
indicate the thicknesses of the individual layers in nm, with
lateral dimensions of 2.92~mm~$\times$~20~$\mu$m. The
heterostructure was prepared by optical lithography, sputter
deposition, and lift-off on a GaAs substrate. Subsequently we
prepared Ag electrodes in contact with the Pt-layer for the voltage
measurements, covered the whole structure with 100-nm thick MgO (for
{\em dc} insulation between the heterostructure and waveguide), and
defined on top a 30-$\mu$m wide and 200-nm thick Au coplanar
waveguide. Two control samples were prepared: a 15~nm Py/15~nm Pt
sample without the 3~nm MgO tunnel-barrier and a 15Py sample without
the normal metal.

\begin{figure}%
  \includegraphics[width=6.5cm, bb=12 10 100 200]{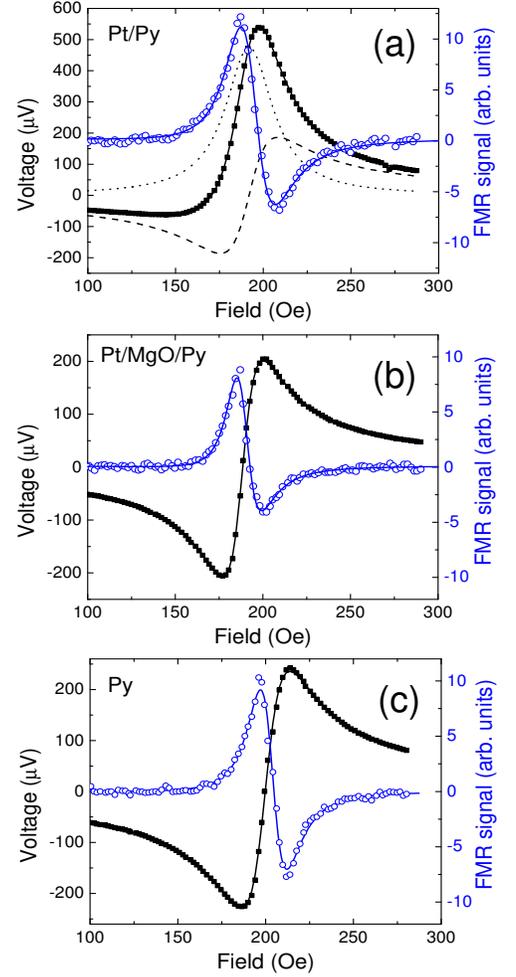}
  \caption{(Color online) FMR derivative spectra are shown (blue open symbols) for (a) Py/Pt, (b) Py/MgO/Pt, and (c) Py samples. Solid lines are fits to a Lorentzian FMR absorption function. The voltage measured along the samples \textit{vs.} field $H_{dc}$ is shown with black solid symbols. Black dotted and dashed lines are fits to Eqs.~(\ref{AMR}) and (\ref{ISHE}), respectively in (a). The solid line in (a) shows the combined fit for the Py/Pt sample. Solid lines in (b) and (c) represent fits of the measured voltage to Eq.~(\ref{AMR}) only.} %
  \label{FMRSHE}%
\end{figure}

The ferromagnetic resonance (FMR) was excited at 4-GHz {\em rf} with
100-mW power, while applying a {\em dc} magnetic field
$\vec{H}_{dc}$ at $\alpha = 45^\circ$ with respect to the waveguide.
The FMR signal was determined from the impedance of the waveguide
\cite{Mosendz-JAP08}; simultaneously the {\em dc} voltage was
measured as a function of $\vec{H}_{dc}$. This is shown in
Fig.~\ref{FMRSHE} for Py/Pt, Py/MgO/Pt and Py.  The FMR peak
positions for all samples are similar and are described by the
Kittel formula with a saturation magnetization for Py$M_{s} = 851$~G
(see Fig.~\ref{FMRSHE}). The FMR linewidths (half width at half
maximum) extracted from fitting to Lorentzian absorption functions
are $\Delta H_{Pt/Py} = 16.9$~Oe for Py/Pt, $\Delta H_{Pt/MgO/Py} =
12.3$~Oe for Py/MgO/Pt, and $\Delta H_{Py} = 12.7$~Oe for Py.  The
difference in FMR linewidths for the Pt/Py and Py samples can be
attributed to the loss of spin momentum in Py due to the relaxation
of the spin accumulation in Pt~\cite{Urban-PRL01}. The linewidth for
the Py/MgO/Pt sample is close to $\Delta H_{Py}$, which already
suggests that spin-pumping is suppressed by the tunnel-barrier, and
non-local damping does not influence the Py layer in the Py/MgO/Pt
structure.

The {\em dc} voltage measured along the samples is shown with solid
symbols in Fig.~\ref{FMRSHE}. We observe a resonant increase of the
{\em dc} voltage along the sample at the FMR position. The signal
measured for Pt/Py has two contributions: (i) anisotropic
magnetoresistance (AMR) and (ii) ISHE~\cite{Mosendz-PRL09}. The
voltage due to AMR is:
\begin{equation}\label{AMR}
  V_{AMR}=I_{rf}
  \Delta R_{AMR}\frac{\sin(2\theta)}{2}\frac{\sin(2\alpha)}{2}\cos\varphi_{0} \; ,
\end{equation}
where $I_{rf}$ is the {\em rf} current, which flows through Py,
$\Delta R_{AMR}$ is the AMR in Py, $\varphi_{0}$ is the phase angle
between magnetization precession and driving {\em rf} field, and
$\theta$ is the cone angle of precession. The voltage due to ISHE
is:
\begin{equation}\label{ISHE}
  V_{ISH}=-\frac{\gamma g_{\uparrow\downarrow} eL \lambda_{sd} \omega}{2\pi \sigma_{Pt} t_{Pt}}
  \sin\alpha\sin^{2}\theta\tanh\left(\frac{t_{Pt}}{2\lambda_{sd}}\right)
  \; ,
\end{equation}
where $g_{\uparrow\downarrow}$ is the spin mixing conductance, $e$
is the electron charge, $L$ is sample length, $\lambda_{sd}$ is the
spin diffusion length in Pt, $\omega$ is the FMR frequency,
$\sigma_{Pt}$ is Pt conductivity, and $t_{Pt}$ is the thickness of
the Pt layer.

\begin{figure*}%
  \includegraphics[scale=2.0,bb=20 18 268 85]{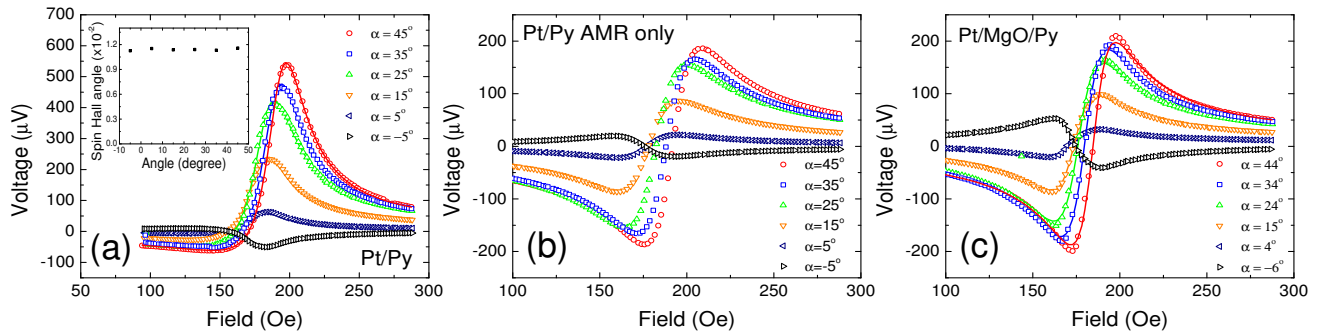}
  \caption{(Color online) Dependence of the measured voltage on the angle $\alpha$ of the applied field with respect to the waveguide axis. (a) Total measured voltage for the Pt/Py sample. The inset shows values of the spin Hall angle as a function of $\alpha$. (b) AMR contribution of the measured voltage for the Pt/Py sample and (c) total measured voltage for the Pt/MgO/Py sample. Note that signals in (b) and (c) are similar.}
  \label{angle}%
\end{figure*}
As shown in Fig.~\ref{FMRSHE}(a), we used Eq.~(\ref{AMR}) for the
AMR contribution (dashed line) and Eq.~(\ref{ISHE}) for the ISHE
contribution (dotted line) to fit the voltage measured for the Py/Pt
sample.  By using a literature value of $\lambda_{sd} = 10$~nm for
Pt \cite{Kurt-APL02}, the only remaining adjustable parameters are
the {\em rf} driving field $h_{rf} = 4.2$~Oe and the spin Hall angle
for Pt $\gamma = 0.0115 \pm 0.0003$ \cite{Mosendz-PRL09}. In
contrast, the single layer Py sample, which is not affected by
spin-pumping, shows a voltage signal, which is described only by the
AMR part. A similar signal is observed for the Py/MgO/Pt sample,
where the insulating tunnel-barrier is present. The relative
strengths of the ISHE and AMR signals depends on the orientation of
the applied field with respect to the waveguide. In
Figs.~\ref{angle}(a) and (c) we show experimental data for several
angles of the applied field for the Py/Pt and Py/MgO/Pt samples. The
measured voltage is consistent with the theoretical prediction, and
results in a constant value of the spin Hall angle in Pt, see inset
in Fig.~\ref{angle}(a). AMR and ISHE contributions of the measured
voltage can be separated, when the data is fitted to the theoretical
model described by Eqs.~(\ref{AMR}) and~(\ref{ISHE}). In
Fig.~\ref{angle}(b) we plot the AMR contribution of the signal
measured in the Py/Pt sample. The AMR contribution of the Py/Pt
sample is the same as the total voltage measured along the Py/MgO/Pt
sample, shown in Fig.~\ref{angle}(c), which indicates that there is
no ISHE on the Py/MgO/Pt sample and consequently there is no spin
current in the Pt layer.  Our results show unambiguously that
spin-pumping is suppressed when a 3-nm thick MgO tunnel-barrier is
inserted at the Py/Pt interface.

In conclusion, we performed FMR with simultaneous transverse voltage
measurements on Py/Pt and Py/MgO/Pt samples. We showed that in the
case of Py/Pt, where spin-pumping is present, a transverse voltage
has contributions from anisotropic magnetoresistance and inverse
spin Hall effect. The transverse voltage for the Py/MgO/Pt sample
has only an anisotropic magnetoresistance contribution. FMR studies
showed that non local damping does not affect the Py layer in the
structure with a tunnel barrier in contrast with the transparent
interface. This result confirms earlier theoretical predictions that
a tunnel-barrier suppresses spin-pumping. The phenomenon, observed
in Ref. \onlinecite{Moriyama-PRL08} appears qualitatively similar to
the predictions of the spin-pumping formalism, however the
spin-pumping detection scheme utilized in the Moriyama's experiment
may be sensitive to a non-equilibrium spin accumulation inside the
ferromagnetic layer itself, caused potentially by structural
imperfections. In this case the tunnel barrier will act as a
non-intrusive probe of the spin-splitting of the chemical potentials
in the ferromagnet induced by the magnetization dynamics
\cite{Tserkovnyak-PRB08}. To this end our studies show that tunnel
barriers inserted at the F/M interface will not be useful for
amplification of spin pumping.

We thank G. Mihajlovi{\' c} and G.E.W. Bauer for valuable
discussions. This work was supported by the U.S. Department of
Energy, Office of Science, Basic Energy Sciences under contract No.
DE-AC02-06CH11357.


\end{document}